\documentclass[superscriptaddress,preprint,nofootinbib,amsmath,amssymb,aps,floatfix,showpacs]{revtex4-1}
\usepackage{graphicx}
\usepackage{bm}
\usepackage{hyperref}
\usepackage{physics}

\begin{document}

\preprint{UT-HET-135}
\preprint{EPHOU-21-013}

\title{Dark Radiation in Spectator Axion-Gauge Models}

\author{Mitsuru Kakizaki}
\email{kakizaki@sci.u-toyama.ac.jp}
\affiliation{Department of Physics, University of Toyama, 3190 Gofuku, Toyama 930-8555, Japan}
\author{Masahito Ogata}
\email{ogata@jodo.sci.u-toyama.ac.jp}
\affiliation{Department of Physics, University of Toyama, 3190 Gofuku, Toyama 930-8555, Japan}
\author{Osamu Seto}
\email{seto@particle.sci.hokudai.ac.jp}
\affiliation{Institute for the Advancement of Higher Education, Hokkaido University, Sapporo 060-0817, Japan}
\affiliation{Department of Physics, Hokkaido University, Sapporo 060-0810, Japan}

\date{March 11, 2022}

\begin{abstract}
In the framework of axion-gauge fields models, primordial gravitational wave perturbations could be generated during the inflationary epoch from not only the quantum fluctuation of gravitons but also the dynamics of hidden gauge fields coupled with an axion field.
We investigate the evolution of the axion and the gauge field of an additional hidden $SU(2)$ gauge group and those energy densities during and after the inflation.
We show that the extra radiation component of the hidden gauge bosons produced by the axion decay can be sizable in the cases where the gauge fields sourced additional gravitational waves is subdominant.
We point out that future measurements of the dark radiation energy, such as CMB-S4, can impose significant constraints on this cosmological scenario.
\end{abstract}

\maketitle

\section{Introduction}

The paradigm of cosmic inflation is successful in resolving the flatness, horizon and monopole problems in the standard Big Bang cosmology~\cite{Starobinsky:1980te,Sato:1980yn,Guth:1980zm}, and has intriguing predictions about primordial perturbations~\cite{Mukhanov:1981xt,Hawking:1982cz,Starobinsky:1982ee,Guth:1982ec}.
One of the important predictions of the inflation is the production of primordial gravitational waves (GW) resulting in the B-mode polarization in the cosmic microwave background (CMB) anisotropy~\cite{Kamionkowski:1996zd,Seljak:1996gy}.
The amplitude of tensor perturbation is usually parametrized by the tensor-to-scalar ratio $r$.
The current upper bound $r<0.056$ at $k=0.002~\mathrm{Mpc}^{-1}$ is obtained at the 95\% confidence level by the Planck and BICEP2/Keck Array data~\cite{Planck:2018jri}.
In the future, the upper limit of the order of $r \lesssim \mathcal{O}(10^{-3})$ 
will be achieved at planned experiments, such as CMB-S4~\cite{Abazajian:2019eic}.
Since the amplitude of tensor perturbation generated by vacuum quantum fluctuation depends only on the Hubble parameter during inflation $H_{\rm inf}^{}$ in a conventional inflation model, this implies that measurement of $r$ can fix the inflationary scale and discriminate inflation models.
However, it should be emphasized that this relationship does not hold in models where gravitational waves are significantly produced by other fields in the early Universe.
For example, in models of axionic inflaton coupled to gauge field~\cite{Anber:2009ua}, tensor perturbations are sourced by the anisotropic stress of vector fields coupled~\cite{Barnaby:2010vf,Sorbo:2011rz}.
For an observed $r$, the inflationary scale is more or less lowered and cannot be determined solely in such models.

The gauge field which an axionic inflaton couples with was extended to be non-Abelian hidden $SU(2)$ gauge field in Chromo-Natural Inflation~\cite{Adshead:2012kp}.
Although it turns out that the original model is hardly compatible with the observational constraints~\cite{Adshead:2013qp,Adshead:2013nka}, 
an inflationary model with a spectator axion and hidden $SU(2)$ gauge fields is a viable variant~\cite{Dimastrogiovanni:2016fuu}.
An efficient gravitational wave production in the linear order of vector perturbation can 
be achieved with the background gauge field, because the hidden $SU(2)$ gauge fields can develop a nontrivial gauge field configuration consistent with homogeneity and isotropy~\cite{Maleknejad:2011jw,Maleknejad:2011sq}.

Although the scenario where the amplitude of gravitational waves sourced by the axion and $SU(2)$ gauge field dynamics in an extremely low scale inflation becomes hierarchically larger than that from the vacuum~\cite{Fujita:2017jwq} is confronted with the constraint of non-Gaussianity~\cite{Papageorgiou:2019ecb},
the gravitational wave spectrum at the CMB scale from the $SU(2)$ gauge field can be of the order of that from the vacuum.
In such cases, $SU(2)$ gauge fields can be copiously produced by axion decays after the inflation, and contribute to the dark radiation energy density, which has been constrained
by CMB measurements.

In this paper, we consider a late time evolution of axions and gauge fields in the axion-$SU(2)$ model~\cite{Dimastrogiovanni:2016fuu} where a fair portion of the primordial gravitational waves is sourced by additional gauge fields.
After the amplitude of the $SU(2)$ gauge bosons is damped, the $SU(2)$ gauge symmetry is left unbroken in the model of Ref.~\cite{Dimastrogiovanni:2016fuu}, without introducing any Higgs field. 
Since the $SU(2)$ gauge bosons remain massless, those gauge bosons produced by the decay of the axion behave as dark radiation and would affect the post inflationary evolution of our Universe. 
We evaluate the abundance of the $SU(2)$ gauge boson dark radiation by solving the evolution of the energy densities of the axion and $SU(2)$ gauge fields, 
and examine the testability of the axion-$SU(2)$ model through measurements of the extra radiation energy density at planned CMB experiments, such as CMB-S4~\cite{Abazajian:2019eic}.
If the two sources of the primordial gravitational waves can be disentangled
at future observations, the parameter space of the model will be significantly narrowed down
in synergy with the dark radiation measurements.

This paper is organized as follows.
In Sec.~\ref{sec:model}, the axion-$SU(2)$ model is briefly reviewed.
Including the effect of the axion decay, we investigate
the evolution of the axion and $SU(2)$ field during and after inflation in Sec.~\ref{sec:evolution},
Various cosmological constraints on this model are discussed in Sec.~\ref{sec:constraints}.
In Sec.~\ref{sec:Neff}, we show the testability of the viable axion-$SU(2)$ model 
in the light of dark radiation.
Section~\ref{sec:summary} is devoted to a summary.

\section{\label{sec:model}Model}

First, we briefly review the axion-$SU(2)$ model~\cite{Dimastrogiovanni:2012ew,Dimastrogiovanni:2016fuu,Fujita:2017jwq}.
In addition to the SM particle contents, we introduce an inflaton $\phi$, a
pseudo-scalar field that we refer to as axion $\chi$,
and a hidden $SU(2)$ gauge field $A_\mu^a$.
We adopt General Relativity with a cosmological constant as gravitational theory,
and assume that the SM Lagrangian remains intact under the presence of interaction between SM fields and the inflaton for reheating processes. As our main results are not dependent on the detail of the inflation model, we are not going to specify or construct an inflation model.
The bottom line is that the inflaton $\phi$ basically accounts for the observed curvature perturbations.
The axion and hidden $SU(2)$ part of the Lagrangian is given by
\begin{align}
\mathcal{L}_{\chi A}^{}=\frac{1}{2}\partial_\mu\chi\partial^\mu\chi-U(\chi) 
 - \frac{1}{4}F_{\mu\nu}^a F^{a \mu\nu}
 + \frac{\lambda}{4f}\chi F_{\mu\nu}^a \widetilde{F}^{a\mu\nu},
\end{align}
with the axion potential
\begin{align}
\label{AG_chipote}
U(\chi)=\mu^4\qty[1+\cos(\frac{\chi}{f})], 
\end{align}
where $F_{\mu\nu}^a = \partial_\mu^{} A_\nu^a - \partial_\nu^{} A_\mu^a
- g \epsilon^{abc} A_\mu^b A_\nu^c$ is the field strength tensor of $A_\mu^a$
with $g$ being the hidden $SU(2)$ coupling constant,
and $\widetilde{F}^{a\mu\nu}=\epsilon^{\mu\nu\lambda\rho}F^a_{\lambda\rho}/2$ is its dual tensor with $\epsilon^{0123}=1/\sqrt{-g}$ for the Levi-Civita tensor.
Here, $\mu$ and $f$ are parameters with mass dimension one, and $\lambda$ is
a dimensionless parameter.

Let us first consider background configurations for the axion and $SU(2)$ gauge field
in the Friedmann-Lemaitre-Robertson-Walker metric.
The homogeneous background component of the axion is expressed by $\chi(t)$.
The background components of the $SU(2)$ gauge field approach
an attractor solution~\cite{Adshead:2012kp, Adshead:2013nka},
\begin{align}
A^a_0=0 ,\quad A^a_i=a(t)Q(t)\delta^a_i, 
\end{align}
which is consistent with the homogeneous and isotropic expansion with the scale factor $a(t)$.
Then, the background field strength is obtained as
\begin{align}
\label{AG_strength}
F^a_{00}=0 ,\quad F^a_{i0}=-F^a_{0i}=a\qty(\dot{Q}+HQ)\delta^a_i ,\quad F^a_{ij}=-g a^2 Q^2 \epsilon^{a}{}_{ij}^{}.
\end{align}
The energy density and pressure of the axion are given by
\begin{align}
\rho_\chi^{} = \frac{1}{2}\dot{\chi}^2+U(\chi), \quad
p_\chi^{} = \frac{1}{2}\dot{\chi}^2-U(\chi),
\end{align}
 and those of the $SU(2)$ gauge field are given by
\begin{align}
\rho_A^{} = \frac{3}{2}\qty[\qty(\dot{Q}+HQ)^2+g^2Q^4], \quad 
p_A^{} = \frac{1}{2}\qty[\qty(\dot{Q}+HQ)^2+g^2Q^4].
\end{align}
We introduce the slow-roll parameter for the axion as
\begin{align}
    \epsilon_\chi = \frac{\dot{\chi}^2}{2 M_{\rm Pl}^2 H^2},
\end{align}
and those for the $SU(2)$ gauge field as
\begin{align}
    \epsilon_A^{} =\epsilon_E^{} +\epsilon_B^{},
    \label{AG_epsilonEB}
\end{align}
and
\begin{align}
    \epsilon_E^{} = \frac{(\dot{Q}+HQ)^2}{M_{\rm{Pl}}^2 H^2},\quad
    \epsilon_B^{} = \frac{g^2 Q^4}{M_{\rm{Pl}}^2 H^2},
\end{align}
with $M_{\rm{Pl}}^{}=2.4 \times 10^{18}~\mathrm{GeV}$ being the reduced Planck scale.
The inflaton energy density $\rho_\phi^{}$ and its slow-roll parameter 
\begin{align}
    \epsilon_\phi^{} = - \frac{\dot{\rho}_\phi}{6 M_{\rm{Pl}}^2 H^3}
\end{align}
depends on models of inflation.
Then, from the Einstein equations at the background, we obtain
\begin{align}
    \label{AG_einsteineq1}
    3 M_{\rm{Pl}}^2 H^2=\rho_\phi + \rho_\chi^{} + \rho_A^{} , \\
    \label{AG_einsteineq2}
    \epsilon_H^{} \equiv -\frac{\dot{H}}{H^2}=\epsilon_\phi + \epsilon_\chi^{} + \epsilon_A^{} .
\end{align}
The equations of motion for the axion $\chi(t)$ and the normalized gauge field $Q(t)$ are given by
\begin{align}
    \label{AG_axionBeq}
    \ddot{\chi}+3H\dot{\chi}+U_{,\chi}+3g\frac{\lambda}{f}Q^2\qty(\dot{Q}+HQ)&=0,\\
    \label{AG_gaugeBeq}
    \ddot{Q}+3H\dot{Q}+\qty(\dot{H}+2H^2)Q+2g^2Q^3-g\frac{\lambda}{f}\dot{\chi}Q^2&=0,
\end{align}
with $U_{,\chi}^{}= d U(\chi)/ d \chi$.

Let us consider the dynamics and evolution of fields at the strong coupling region~\cite{Dimastrogiovanni:2012ew}
\begin{align}
    \label{AG_effpoapp}
    \left(\frac{\lambda Q}{f}\right)^2\gg \frac{3}{{m_Q}^2} ,\quad \left(\frac{\lambda Q}{f}\right)^2\gg 2, 
\end{align}
with $m_Q = g Q/H$ for the initial condition of the gauge field.
During inflation when the following slow-roll approximation
\begin{align}
|\dot{H}|\ll H^2 ,\quad|\ddot{\chi}|\ll|3H\dot{\chi}|, \quad|\ddot{Q}|\ll|3H\dot{Q}|,
\end{align}
 are satisfied, the equations of motion are approximately 
given by~\cite{Dimastrogiovanni:2012ew,Dimastrogiovanni:2016fuu}
\begin{align}
    \label{AG_chieq}
    \frac{1}{H f}\dot{\chi} \simeq& \frac{1}{\lambda}\left( 
    \frac{2 g Q}{H} - \frac{H}{g Q}-
    \frac{f U_{,\chi}^{}}{g^2 \lambda Q^4} \right), \\
    \label{AG_Qeq}
    \frac{1}{H}\dot{Q}\simeq& -Q-\frac{f U_{,\chi}^{}}{3 g \lambda H Q^2} .
\end{align}
The gauge field develops a non-zero minimum field value,
\begin{align}
    \label{AG_agqppro}
    Q_{\rm min}^{}(\chi)=\left(-\frac{f U_{,\chi}^{}}{3g\lambda H} \right)^{1/3} .
\end{align}
At this minimum, the equation of motion for the axion, Eq.~\eqref{AG_chieq}, is rewritten as
\begin{align}
\label{AG_AxionBappeq}
 \frac{1}{H f}\dot{ \chi} = \frac{2}{\lambda} \left( m_{Q, \rm min}^{}(\chi)+\frac{1}{m_{Q, \rm min}^{}(\chi)} \right) ,
\end{align}
with $m_{Q,{\rm min}}(\chi) =  g  Q_{\rm min}(\chi)/H$, 
which will be simply expressed as $m_Q$ in the rest of this paper.

The most interesting aspect of this class of model is the generation of GW background sourced by the gauge field, which would be comparable to the tensor perturbations resulting from the quantum vacuum fluctuation. 
We introduce the parameter $\mathcal{R}_{\rm GW}^{}$ as the ratio of
the power spectrum of gravitational waves sourced by the gauge field 
$\mathcal{P}^{(s)}_h$ to that from the vacuum $\mathcal {P}^{({\rm vac})}_h$:
\begin{align}
\mathcal{R}_{\rm GW}\equiv \frac{\mathcal{P}^{({\rm s})}_h}{\mathcal{P}^{({\rm vac})}_h}.
\end{align}
This ratio is approximately given by~\cite{Dimastrogiovanni:2016fuu,Fujita:2017jwq}
\begin{align}
\mathcal{R}_{\rm GW}\simeq\frac{\epsilon_B^{}}{2} e^{3.6m_Q},
\end{align}
with
\begin{align}
    \epsilon_B^{}= \frac{\pi^2 m_Q^4}{2 g^2} r_{\rm vac} \mathcal{P}_\zeta^{},
\end{align}
where $r_{\rm vac} $ is the vacuum-driven tensor-to-scalar ratio. 

\section{\label{sec:evolution}Evolution of the fields}

During inflation when fields have been within the strong coupling region of Eq.~\eqref{AG_effpoapp}, the axion and gauge fields evolves with taking large field values 
as in Eqs.~\eqref{AG_agqppro} and \eqref{AG_AxionBappeq}, and those energy densities contribute to the total energy density of the Universe only a little.

However, if the energy densities in the axion and gauge fields are not adequately reduced after inflation, the successful predictions about the light element abundances and the CMB power spectrum would be spoiled.
In this section, we follow the evolution of the axion and gauge fields 
after the condition~\eqref{AG_effpoapp} is violated.

When the slow roll condition for the inflation $\phi$ is violated, $\epsilon_{\phi}$ becomes of the order of unity and inflation terminates. For simplicity, we assume the instantaneous reheating where the inflaton decay right after the inflation and the products are instantaneously thermalized.
We here consider the case where the condition~\eqref{AG_effpoapp} has maintained during inflation.
Although the region where $\mathcal{R}_{\rm GW}^{}>1$ is of interest, 
to avoid the non-Gaussianity constraint,
the condition~\eqref{AG_effpoapp} has to be violated during inflation.
In such a case, however, the resultant energy densities of the axion and
gauge fields are extremely diluted by inflation, 
so that no phenomenological signatures are expected.
In the rest of the paper, 
we focus on the region where $0.1<\mathcal{R}_{\rm GW}<1$ to avoid the non-Gaussianity bound, 
and discuss the testability of our model.

After the conditions~\eqref{AG_effpoapp} are violated, 
the axion field no longer follows Eq.~(\ref{AG_AxionBappeq})
and starts to oscillate around the minimum of the potential $U(\chi)$.
Since the equation of state for the axion $w_\chi^{}$ oscillates around zero,
the time-averaged energy density of the axion field attenuates as
\begin{align}
    \label{AG_chirhorad}
    \rho_\chi(a) \simeq 
    \rho_{\chi}^{}(a_{\rm osc}^{}) \qty(\frac{a_{\rm osc}^{}}{a})^3,
\end{align}
with $a_{\rm osc}^{}$ being the scale factor when the axion starts to oscillate.
Assuming that the approximation of Eq.~\eqref{AG_agqppro} is valid 
until the axion oscillation begins, 
the initial axion energy density is estimated as
\begin{align}
    \rho_\chi^{}(a_{\rm osc}^{}) \sim 
    \frac{3^{7/2}H^5}{2g\mu^4}\frac{f^3}{\lambda} ,
\end{align}
for $f>f_\mathrm{cr}$, and
\begin{align}
    \label{AG_rho_osc}
    \rho_\chi^{}(a_{\rm osc}^{}) \sim 
    \left( \frac{6 g^2 H}{\mu^2} \frac{f^3}{\lambda^2} \right)^2 ,
\end{align}
for $f<f_\mathrm{cr}$.
The former corresponds to the case where the first condition in Eq.\eqref{AG_effpoapp} is violated earlier than the second one, and the latter to the opposite case.
Here, we define 
\begin{align}
 f_\mathrm{cr} \equiv \sqrt{\frac{3}{4}}\frac{{H}\lambda}{g}.
\end{align}
Until the axion decays, the energy densities of the axion and $SU(2)$ gauge fields decrease as $a^{-3}$ and $a^{-4}$, respectively. 
Therefore, the energy density of the $SU(2)$ gauge boson becomes negligible compared with that produced by the decay of the axion.

It should be noticed that the axion is not a stable on a cosmological timescale, but 
can decay into the $SU(2)$ gauge fields.
The decay rate of the process $\chi \rightarrow A A$ is given by
\begin{align}
\label{AG_decay_axion}
\Gamma =\frac{3m_\chi^3\lambda^2}{64\pi f^2},
\end{align}
where $m_{\chi}$ is the mass of the axion at the minimum of the potential $U(\chi)$ and the factor $3$ accounts for the dimension of the $SU(2)$ adjoint representation.
We also take the sudden decay approximation for the axion, namely, the coherent oscillating axions instantaneously decay into radiations at $\Gamma = H$ during the radiation dominated era.
Then the energy density of the axion is converted 
into that of the $SU(2)$ gauge boson as $\rho_\chi^{}(a_{\rm dec}^{})=\rho_A^{}(a_{\rm dec}^{})$ due to the energy conservation where $a_{\rm dec}$ is the scale factor at the axion  decay time. After the axion decay, the energy density of the $SU(2)$ gauge boson decreases as $a^{-4}$ independently of the other sectors and is given by
\begin{align}
\label{rho_A}
\rho_A^{}(a) = \rho_\chi^{}(a_{\rm dec}^{}) \left( \frac{a_{\rm dec}^{}} {a}\right)^4 .
\end{align}

Since this $SU(2)$ gauge field does not interact with the SM particles,
it can be regarded as dark radiation.
The total energy density in the radiation dominated era is approximately given by
\begin{align}
\rho_{\rm tot} = \rho_\gamma^{} +\frac{7N^{\rm SM}_{\rm eff}}{4}\frac{\pi^2}{30}T^4_\nu+\rho_d^{},
\end{align}
where $\rho_\gamma^{}$ is the photon energy density, 
$N_{\rm eff}^{\rm SM} \simeq 3.046$ is the effective number of neutrino species in the SM~\cite{Mangano:2005cc,Grohs:2015tfy,deSalas:2016ztq},
and $T_\nu^{}$ is the neutrino temperature.
As is conventionally done, the dark radiation energy density $\rho_d^{}$ will be parametrized in terms of the deviation in the effective neutrino number from the SM value 
$\Delta N_{\rm eff}^{}$ as
\begin{align}
\label{AG_effdegnu}
    \rho_d^{}= \Delta N_{\rm eff}^{} \frac{7}{4} \frac{\pi^2}{30} T^4_\nu.
\end{align}

\begin{figure}[t]
    \centering
	\includegraphics[width=10cm]{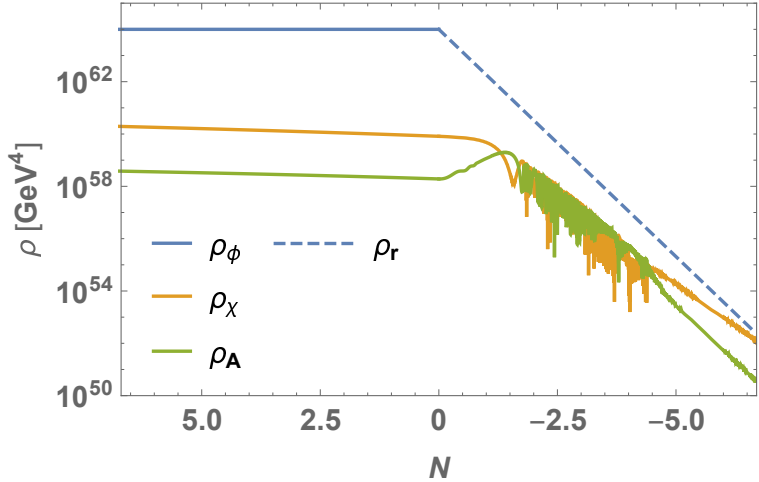}
	\caption{\label{fg:rho_bench}Example of the evolution of the energy density of the inflaton $\rho_\phi^{}$ (blue solid line), that of the axion $\rho_\chi^{}$ (orange solid), and that of the $SU(2)$ gauge boson $\rho_A$ (green solid). 
	For comparison, the energy density of thermalized particles $\rho_r^{}$ is shown (blue dashed). We take  
	$r_{\rm vac}^{} = 10^{-2}$,
	$g=10^{-2}$, $m_{Q*}^{}=2.5$, $f=3.2\times 10^{17}~\mathrm{GeV}$, $\lambda=230$ and
	$\chi_*^{}/f = \pi/2$.}
\end{figure}

Figure~\ref{fg:rho_bench} shows an example of the evolution of the energy density of the inflaton $\rho_\phi^{}$ (blue solid line), that of the axion $\rho_\chi^{}$ (orange solid), that of the $SU(2)$ gauge boson $\rho_A$ (green solid), and that of radiation $\rho_r^{}$ generated by the inflaton decay (blue dashed).
Here, for the sake of convenience, we replace the time coordinate $t$ by the number of e-folds defined by
\begin{align}
\label{Eq:N-t relation}
 N = \ln\left(\frac{a_\mathrm{end}}{a(t)}\right), 
\end{align}
with $a_\mathrm{end}$ being the scale factor at the end of the inflation.
As a benchmark point, we take
\begin{align}
   \label{eq:bmp}
   r_{\rm vac}^{} = 10^{-2}, \quad g=10^{-2}, \quad m_{Q*}^{}=2.5.
\end{align}
The initial condition is set at the pivot scale of CMB observations.
The initial axion phase is set at $\chi_*^{}/f = \pi/2$ so that gravitational waves driven
by the gauge fields is maximized, leading to
\begin{align}
\label{AG_mulambda}
\mu= \left( \frac{3 \lambda H_*^4 m_{Q*}^3}{g^2} \right)^{1/4}.
\end{align}
The axion decay constant and the axion-gauge coupling are set by
\begin{align}
    \label{eq:param}
    f=3.2\times 10^{17}~\mathrm{GeV} ,\quad \lambda=230.
\end{align}
The enhancement of the energy density of the $SU(2)$ gauge boson 
relative to the $a^{-4}$ scaling law at around $N=-2.5$ is attributed the 
coupling with the axion, from which energy is transferred.
The instantaneous axion decay occurs at $N=-6.7$, 
when the energy density of the axion is converted to that of the $SU(2)$ gauge boson.
Since our analysis is centered on the energy density of dark radiation, 
the intrinsic energy density of the $SU(2)$ gauge boson is disregarded 
in the rest of this paper.

\section{\label{sec:constraints}Constraints on the Model}

In this section, we discuss constraints on the model parameter space.
Here, we consider the stability of the background field 
and contributions the scalar perturbations of the CMB.

Since the gauge field generates large tensor perturbations in this model,
strong back reaction to the dynamics of the background field may disturb our assumptions.
In particular, the background gauge field tends to be destabilized by
the back reaction term $\mathcal{T}^Q_{\rm BR}$, which modifies Eq.~\eqref{AG_gaugeBeq}.
The condition that the effect of $\mathcal{T}^Q_{\rm BR}$ is negligible is investigated in Refs.~\cite{Maleknejad:2018nxz, Papageorgiou:2019ecb}, and given by
\begin{align}
 g\ll\qty(\frac{24\pi^2}{2.3e^{3.9m_{Q*}}}\frac{1}{1+m_{Q*}^{-2}})^{1/2},
\end{align}
where the subscript $*$ indicates CMB-scale values.

Next, we consider the constraints imposed by improper normalization of the scalar perturbations.
Using the fact that $\epsilon_B^{}$ dominates in the Hubble slow roll parameter
$\epsilon_H^{}$,
the power spectrum of the primordial curvature perturbations $\mathcal{P}_\zeta$
is found to be~\cite{Papageorgiou:2019ecb}
\begin{align}
\mathcal{P}_\zeta\simeq \left. \frac{1}{2\epsilon_{H}}\qty(\frac{\epsilon_{\phi}}{\epsilon_{H}})\qty(\frac{H}{2\pi M_{\rm Pl}^{}})^2 \right|_*
\simeq
\left. \frac{H^2}{8\pi^2 M_{\rm Pl}^2}\frac{\epsilon_{\phi}}{\qty(\epsilon_{\phi}+\epsilon_{B})^2}
 \right|_*.
\end{align}
The above equation can be rewritten as 
\begin{align}
\mathcal{P}_\zeta
\simeq
\left.
\frac{g^2}{8\pi^2 m_{Q}^4}\frac{\epsilon_{B}/\epsilon_{\phi}}{\qty(1+\epsilon_{B}/\epsilon_{\phi})^2}
\right|_*
> \frac{g^2}{32\pi^2 m_{Q*}^4}.
\end{align}
The last inequality constrains viable values of $g$ and $m_Q^{}$.
In addition, for stabilizing the background fields, the range of $m_Q^{}$ is bounded from the below as $m_Q^{}>\sqrt{2}$~\cite{Dimastrogiovanni:2012ew}.

\begin{figure}[t]
    \centering
	\includegraphics[width=0.7\textwidth]{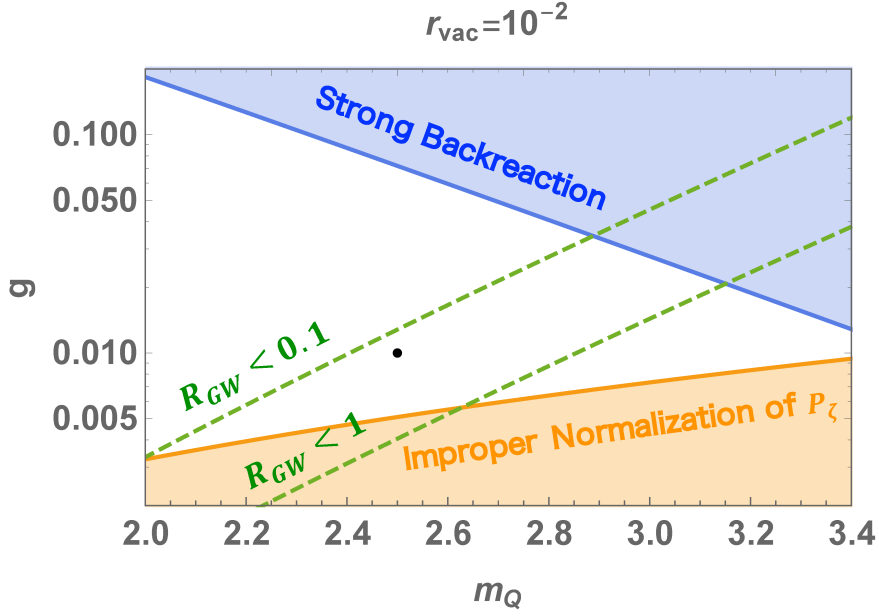}
	\caption{\label{fg:gm_const}
	Allowed regions for $r_{\rm vac}=10^{-2}$ in the ($m_Q^{}$,$g$) plain.
    The blue (orange) region is excluded by the condition of strong back reaction
    (improper normalization of $\mathcal{P}_\zeta^{}$).
    The green dashed lines denote the contours of $\mathcal{R}_{\rm GW}^{}$.
    The blob indicates the benchmark point.}
\end{figure}

Figure~\ref{fg:gm_const} shows allowed regions in the ($m_Q^{}$,$g$) plain.
The left (right) panel corresponds with $r_{\rm vac}=0.04$ ($r_{\rm vac}=10^{-2}$).
The blue (orange) region is excluded by the condition of strong back reaction
(improper normalization of $\mathcal{P}_\zeta^{}$).
The green dashed lines denote the contours of $\mathcal{R}_{\rm GW}^{}$.
In the upper right region, $\mathcal{R}_{\rm GW}^{}<1$.
This figure is consistent with the results presented in Ref.~\cite{Papageorgiou:2019ecb}.
The blob indicates the benchmark point, Eq~\eqref{eq:bmp}.

One of the most stringent constraints is imposed by non-observation of
primordial non-Gaussianity~\cite{Planck:2015zfm}.
Although the tensor mode of the $SU(2)$ gauge field does not couple to the scalar one at the linear order, 
it is pointed out that it can significantly contribute to the scalar perturbations at the quadratic order~\cite{Papageorgiou:2019ecb}.
This large quadratic order contribution is caused by 
loop diagrams containing the enhanced tensor mode,
and leads to highly non-Gaussian curvature perturbation.
To be consistent with the observation,
the nonlinear contribution to the power spectrum $\delta \mathcal{P}_\phi^{}$ needs to be sufficiently small 
compared to the linear one $\mathcal{P}_\phi^{}$.
A detailed computation of the bispectrum in this model is required for the accurate bounds on the parameters.
By rule of thumb, the reported bound on the equilateral
non-Gaussianity, $f_{\rm NL}^{\rm equil} \lesssim \mathcal{O}(100)$,
is converted into the upper bound on the ratio~\cite{Papageorgiou:2019ecb}
\begin{align}
\label{AG_nongau1}
\mathcal{R}_{\delta\phi}^{} \equiv \frac{\delta\mathcal{P}_\phi}{\mathcal{P}_\phi}<0.1.
\end{align}
In the range of $2.5 <m_Q <3.5$, 
the numerical results about $\mathcal{R}_{\delta\phi}^{}$ are
accurately fitted by the function~\cite{Papageorgiou:2019ecb}
\begin{align}
\label{AG_nongau2}
\mathcal{R}_{\delta\phi}^{} \simeq \frac{5\times 10^{-12}}{\qty(1+\epsilon_B/\epsilon_\phi)^2}
m_Q^{11} e^{7m_Q} N_k^2 r_{\rm vac}^2,
\end{align}
where $N_k^{}$ denotes the number of e-folds during which the axion rolls from the CMB scale until the potential minimum.

\begin{figure}[t]
    \centering
	\includegraphics[width=0.49\textwidth]{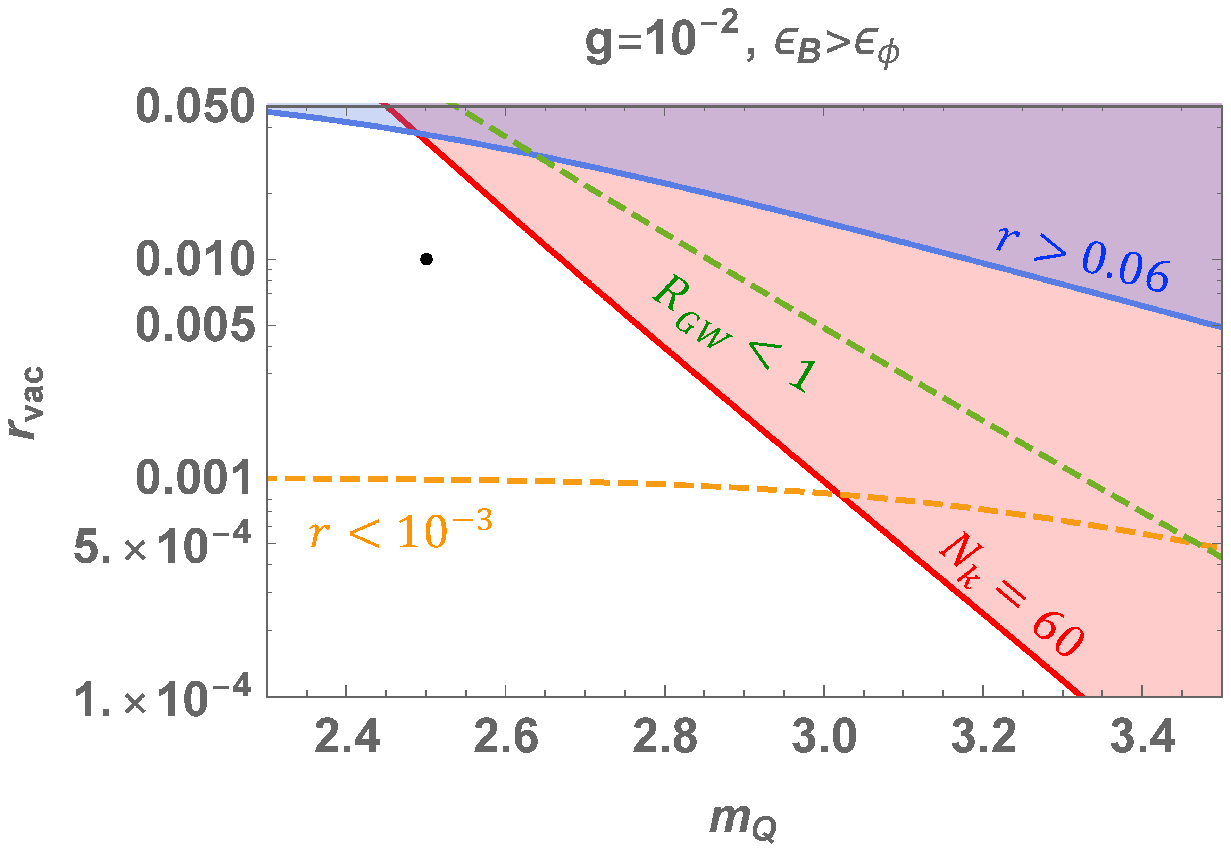}
	\includegraphics[width=0.49\textwidth]{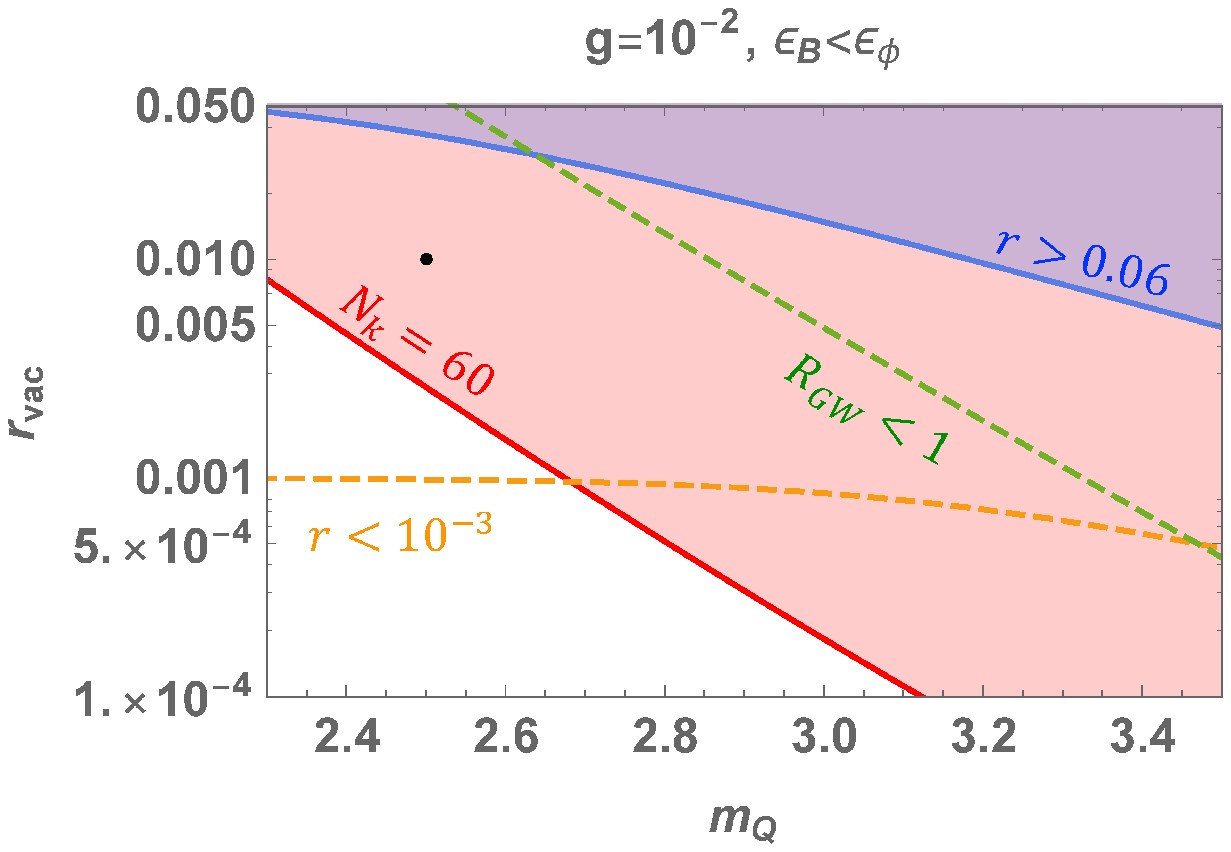}
	\caption{\label{fg:ng_const}Allowed regions in the ($m_Q^{}$, $r_{\rm vac}$) plain for $g=10^{-2}$.
	For details, see the text.}
	\label{Fig:nonG}
\end{figure}

Imposing the non-Gaussianity bound, in Fig.~\ref{Fig:nonG}, 
we show allowed regions in the ($m_Q^{}$,$r_{\rm vac}$) plain for $g=10^{-2}$.
The left (right) panel is for $\epsilon_B^{}>\epsilon_\phi^{}$ ($\epsilon_B^{}<\epsilon_\phi^{}$).
The red regions above the red solid lines are excluded by non-Gaussianity.
The red solid lines are contours of $N_k^{}$.
The blue solid line is the current upper bound on the tensor-to-scalar ratio $r=0.06$ while the orange dashed line shows the future sensitivity expected at CMB-S4 $r=10^{-3}$.
The green dashed line corresponds to $\mathcal{R}_{\rm GW}^{}=1$.
This figure is consistent with the results presented in Ref.~\cite{Papageorgiou:2019ecb}.
It turns out that there is a viable parameter space where
the gravitational waves sourced by the $SU(2)$ gauge field 
can be as strong as those from the vacuum
while avoiding the non-Gaussianity constraint.
Our benchmark point~(\ref{eq:bmp}) is free from the non-Gaussianity constraint for $\epsilon_B^{}> \epsilon_\phi^{}$ and has cosmological history where the axion starts to oscillate after inflation, $a_\mathrm{osc} > a_\mathrm{end}$.
For a $\epsilon_B^{}> \epsilon_\phi^{}$ case, a smaller $m_Q^{}$ is needed to realize $a_\mathrm{osc}^{} > a_\mathrm{end}^{}$.
In the next section, we will only consider our benchmark point~(\ref{eq:bmp}) with $\epsilon_B^{}> \epsilon_\phi^{}$.

The spectator axion does not contribute to the curvature perturbations,
since the axion is massive during the inflationary epoch 
and its perturbations damp at the superhorizon scale.
The motion of the axion is not an ordinary slow roll caused 
by the Hubble friction on the flat potential, 
but actually an adiabatic trapped roll by the strong coupling.

\section{\label{sec:Neff}Energy density of dark radiation}

Finally, we examine to what extent the energy density of dark radiation is left
within the limits presented in the previous section,
and discuss a future prospect for detecting the energy density of dark radiation.

In the cases where the contribution from the gauge field to the gravitational wave power spectrum dominates over that from the vacuum $\mathcal{R}_{\rm GW}^{}>1$,
the ratio of the non-linear contribution to the curvature perturbation $\mathcal{R}_{\delta \phi}^{}$ is large so that the model is tightly constrained by the non-Gaussianity bound~\cite{Papageorgiou:2019ecb}.

\begin{figure}[t]
	\begin{center}
	\includegraphics[width=0.7\textwidth]{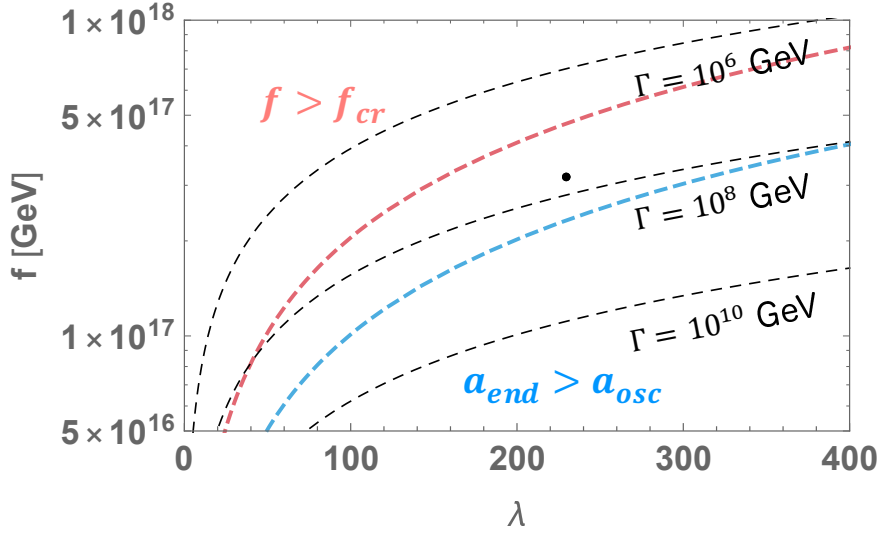}
	\caption{\label{fg:gamma_contour} Contour plot for the axion decay rate $\Gamma$ in the ($\lambda$, $f$) plain for $H_*^{} = 2.4 \times 10^{13}~\mathrm{GeV}$, $g=10^{-2}$ and $m_{Q*}=2.5$.}
	\end{center}
\end{figure}

We focus on the cases where the gauge field substantially contributes to 
the gravitational wave power spectrum such that $0.1 < \mathcal{R}_{\rm GW}^{}<1$,
and estimate the energy density the gauge field.
As shown in the right panel of Fig.~\ref{fg:gm_const}, our benchmark point of Eq.~\eqref{eq:bmp} with $\epsilon_B^{}> \epsilon_\phi^{}$ falls within this range.
Figure~\ref{fg:gamma_contour} shows the contours of the axion decay rate $\Gamma$ in the ($\lambda$, $f$) plain.
The viable axion decay rate is found to be in the range 
$10^{7}~\mathrm{GeV} \lesssim \Gamma \lesssim 10^{9} ~\mathrm{GeV}$ and is realized for an appropriate choice of the model parameters.
The parameter set of Eq.~\eqref{eq:param} is indicated as the blob in Figure~\ref{fg:gamma_contour}.
For the above parameter set, the axion decay rate and energy density of the axion at the decay time are found to be
\begin{align}
   \Gamma=4.7\times10^{7}~\mathrm{GeV} ,\quad 
    \rho_{\chi,{\rm dec}}^{1/4}
    \equiv \left[ \rho_{\chi}^{}(a_{\rm dec}^{}) \right]^{1/4}
    =1.1\times10^{13}~\mathrm{GeV},
\end{align}
respectively.
As a result, the predicted value of the energy density of the gauge field in terms of the effective neutrino number is found to be
\begin{align}
   N_{\rm eff}^{\rm CMB}= 0.066 .
\end{align}

\begin{figure}[t]
	\begin{center}
	\includegraphics[width=0.7\textwidth]{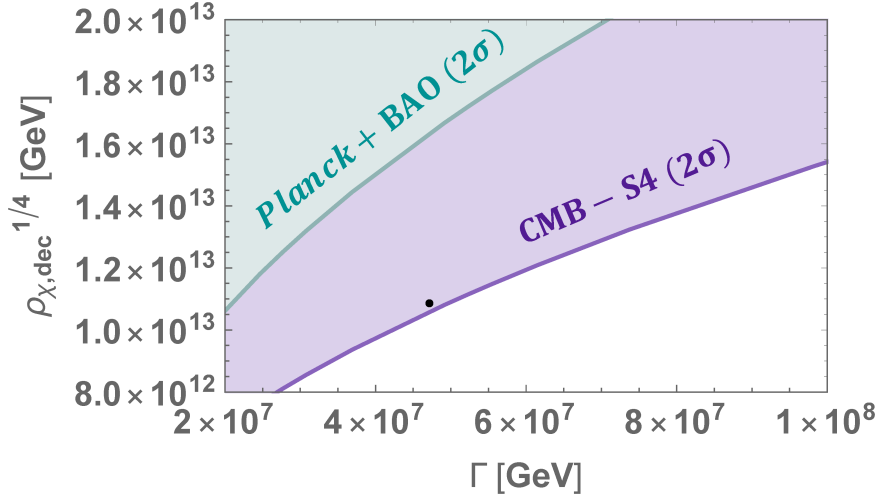}
	\caption{\label{fg:DeltaNeff_contour} Contour plot for $\Delta N_{\rm eff}^{}$ in the ($\Gamma$, $\rho_{\chi,{\rm dec}})$ plain.
	The green region is excluded by the Planck $+$ BAO data at the 2$\sigma$ level.
	The purple region shows a future sensitivity expected at CMB-S4 at the 2$\sigma$ level.}
	\end{center}
\end{figure}

The Planck 2018 data combined with baryon acoustic oscillation data constrain the 
effective extra relativistic degrees of freedom as~\cite{Planck:2018vyg}
\begin{align}
N_{\rm eff}^{\rm CMB}= 2.99 \pm 0.17,
\end{align}
at the $1\sigma$ level.
This result is consistent with the SM prediction 
$N^{\rm SM}_{\rm eff} \approx 3.046$~\cite{Mangano:2005cc,Grohs:2015tfy,deSalas:2016ztq}.
The future CMB-S4 experiment is expected to reach a precision of~\cite{Abazajian:2019eic}
\begin{align}
\Delta N_{\rm eff}^{\rm CMB} < 0.03,
\end{align}
at the $1\sigma$ level.
The contours of the predicted values of $\Delta N_{\rm eff}^{}$ in our model are 
plotted in Fig.~\ref{fg:DeltaNeff_contour}.
The green region is excluded by the Planck $+$ BAO data at the 2$\sigma$ level.
The purple region displays an expected future sensitivity at CMB-S4 at the 2$\sigma$ level.
Therefore, measurements of $N_{\rm eff}^{}$ at future CMB experiments have a detectability of dark radiation originating from the dynamics of the axion and gauge fields and can constrain models with axion-gauge fields.

\section{\label{sec:summary}Summary}

In the axion-gauge fields model, the nontrivial configuration and dynamics of the axion and hidden gauge fields during the cosmic inflation can significantly contribute to  the stochastic gravitational wave background. 
In the cases where this contribution is of the same order of that from the vacuum,  the energy density of the axion is relatively large after the inflation.
Since hidden gauge fields are produced through the axion decay, they contribute to the dark radiation energy density, leading to a larger neutrino effective degrees of freedom $\Delta N_{\rm eff}$ than the SM prediction. 
We have computed $\Delta N_{\rm eff}$ in the model with the axion and hidden $SU(2)$ gauge field in the light of the experimental results of the tensor-to-scalar ratio and the non-gaussianity.
It has been pointed out that precise measurements of $N_{\rm eff}$ at future CMB experiments can test viable parameter regions in this model whose predicted ratio of the tensor fluctuations from the gauge field to those from the vacuum is $\mathcal{O}(10^{-1})$.

\begin{acknowledgments}
The authors would like to thank Tomohiro Fujita for useful discussion and comments.
The work was supported in part by the Japan Society for the Promotion of Science (JSPS) Grant-in-Aid for Scientific Research KAKENHI Grant Numbers No.~20H00160, No.~21K03571 (M.K.), No.~19K03860, No.~19K03865 and No.~21H00060 (O.S.).
\end{acknowledgments}

\bibliographystyle{apsrev4-1}

\bibliography{ref}

\begin{thebibliography}{29}%
\makeatletter
\providecommand \@ifxundefined [1]{%
 \@ifx{#1\undefined}
}%
\providecommand \@ifnum [1]{%
 \ifnum #1\expandafter \@firstoftwo
 \else \expandafter \@secondoftwo
 \fi
}%
\providecommand \@ifx [1]{%
 \ifx #1\expandafter \@firstoftwo
 \else \expandafter \@secondoftwo
 \fi
}%
\providecommand \natexlab [1]{#1}%
\providecommand \enquote  [1]{``#1''}%
\providecommand \bibnamefont  [1]{#1}%
\providecommand \bibfnamefont [1]{#1}%
\providecommand \citenamefont [1]{#1}%
\providecommand \href@noop [0]{\@secondoftwo}%
\providecommand \href [0]{\begingroup \@sanitize@url \@href}%
\providecommand \@href[1]{\@@startlink{#1}\@@href}%
\providecommand \@@href[1]{\endgroup#1\@@endlink}%
\providecommand \@sanitize@url [0]{\catcode `\\12\catcode `\$12\catcode
  `\&12\catcode `\#12\catcode `\^12\catcode `\_12\catcode `\%12\relax}%
\providecommand \@@startlink[1]{}%
\providecommand \@@endlink[0]{}%
\providecommand \url  [0]{\begingroup\@sanitize@url \@url }%
\providecommand \@url [1]{\endgroup\@href {#1}{\urlprefix }}%
\providecommand \urlprefix  [0]{URL }%
\providecommand \Eprint [0]{\href }%
\providecommand \doibase [0]{http://dx.doi.org/}%
\providecommand \selectlanguage [0]{\@gobble}%
\providecommand \bibinfo  [0]{\@secondoftwo}%
\providecommand \bibfield  [0]{\@secondoftwo}%
\providecommand \translation [1]{[#1]}%
\providecommand \BibitemOpen [0]{}%
\providecommand \bibitemStop [0]{}%
\providecommand \bibitemNoStop [0]{.\EOS\space}%
\providecommand \EOS [0]{\spacefactor3000\relax}%
\providecommand \BibitemShut  [1]{\csname bibitem#1\endcsname}%
\let\auto@bib@innerbib\@empty
\bibitem [{\citenamefont {Starobinsky}(1980)}]{Starobinsky:1980te}%
  \BibitemOpen
  \bibfield  {author} {\bibinfo {author} {\bibfnamefont {A.~A.}\ \bibnamefont
  {Starobinsky}},\ }\href {\doibase 10.1016/0370-2693(80)90670-X} {\bibfield
  {journal} {\bibinfo  {journal} {Phys. Lett. B}\ }\textbf {\bibinfo {volume}
  {91}},\ \bibinfo {pages} {99} (\bibinfo {year} {1980})}\BibitemShut {NoStop}%
\bibitem [{\citenamefont {Sato}(1981)}]{Sato:1980yn}%
  \BibitemOpen
  \bibfield  {author} {\bibinfo {author} {\bibfnamefont {K.}~\bibnamefont
  {Sato}},\ }\href@noop {} {\bibfield  {journal} {\bibinfo  {journal} {Mon.
  Not. Roy. Astron. Soc.}\ }\textbf {\bibinfo {volume} {195}},\ \bibinfo
  {pages} {467} (\bibinfo {year} {1981})}\BibitemShut {NoStop}%
\bibitem [{\citenamefont {Guth}(1981)}]{Guth:1980zm}%
  \BibitemOpen
  \bibfield  {author} {\bibinfo {author} {\bibfnamefont {A.~H.}\ \bibnamefont
  {Guth}},\ }\href {\doibase 10.1103/PhysRevD.23.347} {\bibfield  {journal}
  {\bibinfo  {journal} {Phys. Rev. D}\ }\textbf {\bibinfo {volume} {23}},\
  \bibinfo {pages} {347} (\bibinfo {year} {1981})}\BibitemShut {NoStop}%
\bibitem [{\citenamefont {Mukhanov}\ and\ \citenamefont
  {Chibisov}(1981)}]{Mukhanov:1981xt}%
  \BibitemOpen
  \bibfield  {author} {\bibinfo {author} {\bibfnamefont {V.~F.}\ \bibnamefont
  {Mukhanov}}\ and\ \bibinfo {author} {\bibfnamefont {G.~V.}\ \bibnamefont
  {Chibisov}},\ }\href@noop {} {\bibfield  {journal} {\bibinfo  {journal} {JETP
  Lett.}\ }\textbf {\bibinfo {volume} {33}},\ \bibinfo {pages} {532} (\bibinfo
  {year} {1981})}\BibitemShut {NoStop}%
\bibitem [{\citenamefont {Hawking}(1982)}]{Hawking:1982cz}%
  \BibitemOpen
  \bibfield  {author} {\bibinfo {author} {\bibfnamefont {S.~W.}\ \bibnamefont
  {Hawking}},\ }\href {\doibase 10.1016/0370-2693(82)90373-2} {\bibfield
  {journal} {\bibinfo  {journal} {Phys. Lett. B}\ }\textbf {\bibinfo {volume}
  {115}},\ \bibinfo {pages} {295} (\bibinfo {year} {1982})}\BibitemShut
  {NoStop}%
\bibitem [{\citenamefont {Starobinsky}(1982)}]{Starobinsky:1982ee}%
  \BibitemOpen
  \bibfield  {author} {\bibinfo {author} {\bibfnamefont {A.~A.}\ \bibnamefont
  {Starobinsky}},\ }\href {\doibase 10.1016/0370-2693(82)90541-X} {\bibfield
  {journal} {\bibinfo  {journal} {Phys. Lett. B}\ }\textbf {\bibinfo {volume}
  {117}},\ \bibinfo {pages} {175} (\bibinfo {year} {1982})}\BibitemShut
  {NoStop}%
\bibitem [{\citenamefont {Guth}\ and\ \citenamefont {Pi}(1982)}]{Guth:1982ec}%
  \BibitemOpen
  \bibfield  {author} {\bibinfo {author} {\bibfnamefont {A.~H.}\ \bibnamefont
  {Guth}}\ and\ \bibinfo {author} {\bibfnamefont {S.~Y.}\ \bibnamefont {Pi}},\
  }\href {\doibase 10.1103/PhysRevLett.49.1110} {\bibfield  {journal} {\bibinfo
   {journal} {Phys. Rev. Lett.}\ }\textbf {\bibinfo {volume} {49}},\ \bibinfo
  {pages} {1110} (\bibinfo {year} {1982})}\BibitemShut {NoStop}%
\bibitem [{\citenamefont {Kamionkowski}\ \emph {et~al.}(1997)\citenamefont
  {Kamionkowski}, \citenamefont {Kosowsky},\ and\ \citenamefont
  {Stebbins}}]{Kamionkowski:1996zd}%
  \BibitemOpen
  \bibfield  {author} {\bibinfo {author} {\bibfnamefont {M.}~\bibnamefont
  {Kamionkowski}}, \bibinfo {author} {\bibfnamefont {A.}~\bibnamefont
  {Kosowsky}}, \ and\ \bibinfo {author} {\bibfnamefont {A.}~\bibnamefont
  {Stebbins}},\ }\href {\doibase 10.1103/PhysRevLett.78.2058} {\bibfield
  {journal} {\bibinfo  {journal} {Phys. Rev. Lett.}\ }\textbf {\bibinfo
  {volume} {78}},\ \bibinfo {pages} {2058} (\bibinfo {year} {1997})},\ \Eprint
  {http://arxiv.org/abs/astro-ph/9609132} {arXiv:astro-ph/9609132} \BibitemShut
  {NoStop}%
\bibitem [{\citenamefont {Seljak}\ and\ \citenamefont
  {Zaldarriaga}(1997)}]{Seljak:1996gy}%
  \BibitemOpen
  \bibfield  {author} {\bibinfo {author} {\bibfnamefont {U.}~\bibnamefont
  {Seljak}}\ and\ \bibinfo {author} {\bibfnamefont {M.}~\bibnamefont
  {Zaldarriaga}},\ }\href {\doibase 10.1103/PhysRevLett.78.2054} {\bibfield
  {journal} {\bibinfo  {journal} {Phys. Rev. Lett.}\ }\textbf {\bibinfo
  {volume} {78}},\ \bibinfo {pages} {2054} (\bibinfo {year} {1997})},\ \Eprint
  {http://arxiv.org/abs/astro-ph/9609169} {arXiv:astro-ph/9609169} \BibitemShut
  {NoStop}%
\bibitem [{\citenamefont {Akrami}\ \emph {et~al.}(2020)\citenamefont {Akrami}
  \emph {et~al.}}]{Planck:2018jri}%
  \BibitemOpen
  \bibfield  {author} {\bibinfo {author} {\bibfnamefont {Y.}~\bibnamefont
  {Akrami}} \emph {et~al.} (\bibinfo {collaboration} {Planck}),\ }\href
  {\doibase 10.1051/0004-6361/201833887} {\bibfield  {journal} {\bibinfo
  {journal} {Astron. Astrophys.}\ }\textbf {\bibinfo {volume} {641}},\ \bibinfo
  {pages} {A10} (\bibinfo {year} {2020})},\ \Eprint
  {http://arxiv.org/abs/1807.06211} {arXiv:1807.06211 [astro-ph.CO]}
  \BibitemShut {NoStop}%
\bibitem [{\citenamefont {Abazajian}\ \emph {et~al.}(2019)\citenamefont
  {Abazajian} \emph {et~al.}}]{Abazajian:2019eic}%
  \BibitemOpen
  \bibfield  {author} {\bibinfo {author} {\bibfnamefont {K.}~\bibnamefont
  {Abazajian}} \emph {et~al.},\ }\href@noop {} {\  (\bibinfo {year} {2019})},\
  \Eprint {http://arxiv.org/abs/1907.04473} {arXiv:1907.04473 [astro-ph.IM]}
  \BibitemShut {NoStop}%
\bibitem [{\citenamefont {Anber}\ and\ \citenamefont
  {Sorbo}(2010)}]{Anber:2009ua}%
  \BibitemOpen
  \bibfield  {author} {\bibinfo {author} {\bibfnamefont {M.~M.}\ \bibnamefont
  {Anber}}\ and\ \bibinfo {author} {\bibfnamefont {L.}~\bibnamefont {Sorbo}},\
  }\href {\doibase 10.1103/PhysRevD.81.043534} {\bibfield  {journal} {\bibinfo
  {journal} {Phys. Rev. D}\ }\textbf {\bibinfo {volume} {81}},\ \bibinfo
  {pages} {043534} (\bibinfo {year} {2010})},\ \Eprint
  {http://arxiv.org/abs/0908.4089} {arXiv:0908.4089 [hep-th]} \BibitemShut
  {NoStop}%
\bibitem [{\citenamefont {Barnaby}\ and\ \citenamefont
  {Peloso}(2011)}]{Barnaby:2010vf}%
  \BibitemOpen
  \bibfield  {author} {\bibinfo {author} {\bibfnamefont {N.}~\bibnamefont
  {Barnaby}}\ and\ \bibinfo {author} {\bibfnamefont {M.}~\bibnamefont
  {Peloso}},\ }\href {\doibase 10.1103/PhysRevLett.106.181301} {\bibfield
  {journal} {\bibinfo  {journal} {Phys. Rev. Lett.}\ }\textbf {\bibinfo
  {volume} {106}},\ \bibinfo {pages} {181301} (\bibinfo {year} {2011})},\
  \Eprint {http://arxiv.org/abs/1011.1500} {arXiv:1011.1500 [hep-ph]}
  \BibitemShut {NoStop}%
\bibitem [{\citenamefont {Sorbo}(2011)}]{Sorbo:2011rz}%
  \BibitemOpen
  \bibfield  {author} {\bibinfo {author} {\bibfnamefont {L.}~\bibnamefont
  {Sorbo}},\ }\href {\doibase 10.1088/1475-7516/2011/06/003} {\bibfield
  {journal} {\bibinfo  {journal} {JCAP}\ }\textbf {\bibinfo {volume} {06}},\
  \bibinfo {pages} {003} (\bibinfo {year} {2011})},\ \Eprint
  {http://arxiv.org/abs/1101.1525} {arXiv:1101.1525 [astro-ph.CO]} \BibitemShut
  {NoStop}%
\bibitem [{\citenamefont {Adshead}\ and\ \citenamefont
  {Wyman}(2012)}]{Adshead:2012kp}%
  \BibitemOpen
  \bibfield  {author} {\bibinfo {author} {\bibfnamefont {P.}~\bibnamefont
  {Adshead}}\ and\ \bibinfo {author} {\bibfnamefont {M.}~\bibnamefont
  {Wyman}},\ }\href {\doibase 10.1103/PhysRevLett.108.261302} {\bibfield
  {journal} {\bibinfo  {journal} {Phys. Rev. Lett.}\ }\textbf {\bibinfo
  {volume} {108}},\ \bibinfo {pages} {261302} (\bibinfo {year} {2012})},\
  \Eprint {http://arxiv.org/abs/1202.2366} {arXiv:1202.2366 [hep-th]}
  \BibitemShut {NoStop}%
\bibitem [{\citenamefont {Adshead}\ \emph
  {et~al.}(2013{\natexlab{a}})\citenamefont {Adshead}, \citenamefont
  {Martinec},\ and\ \citenamefont {Wyman}}]{Adshead:2013qp}%
  \BibitemOpen
  \bibfield  {author} {\bibinfo {author} {\bibfnamefont {P.}~\bibnamefont
  {Adshead}}, \bibinfo {author} {\bibfnamefont {E.}~\bibnamefont {Martinec}}, \
  and\ \bibinfo {author} {\bibfnamefont {M.}~\bibnamefont {Wyman}},\ }\href
  {\doibase 10.1103/PhysRevD.88.021302} {\bibfield  {journal} {\bibinfo
  {journal} {Phys. Rev. D}\ }\textbf {\bibinfo {volume} {88}},\ \bibinfo
  {pages} {021302} (\bibinfo {year} {2013}{\natexlab{a}})},\ \Eprint
  {http://arxiv.org/abs/1301.2598} {arXiv:1301.2598 [hep-th]} \BibitemShut
  {NoStop}%
\bibitem [{\citenamefont {Adshead}\ \emph
  {et~al.}(2013{\natexlab{b}})\citenamefont {Adshead}, \citenamefont
  {Martinec},\ and\ \citenamefont {Wyman}}]{Adshead:2013nka}%
  \BibitemOpen
  \bibfield  {author} {\bibinfo {author} {\bibfnamefont {P.}~\bibnamefont
  {Adshead}}, \bibinfo {author} {\bibfnamefont {E.}~\bibnamefont {Martinec}}, \
  and\ \bibinfo {author} {\bibfnamefont {M.}~\bibnamefont {Wyman}},\ }\href
  {\doibase 10.1007/JHEP09(2013)087} {\bibfield  {journal} {\bibinfo  {journal}
  {JHEP}\ }\textbf {\bibinfo {volume} {09}},\ \bibinfo {pages} {087} (\bibinfo
  {year} {2013}{\natexlab{b}})},\ \Eprint {http://arxiv.org/abs/1305.2930}
  {arXiv:1305.2930 [hep-th]} \BibitemShut {NoStop}%
\bibitem [{\citenamefont {Dimastrogiovanni}\ \emph {et~al.}(2017)\citenamefont
  {Dimastrogiovanni}, \citenamefont {Fasiello},\ and\ \citenamefont
  {Fujita}}]{Dimastrogiovanni:2016fuu}%
  \BibitemOpen
  \bibfield  {author} {\bibinfo {author} {\bibfnamefont {E.}~\bibnamefont
  {Dimastrogiovanni}}, \bibinfo {author} {\bibfnamefont {M.}~\bibnamefont
  {Fasiello}}, \ and\ \bibinfo {author} {\bibfnamefont {T.}~\bibnamefont
  {Fujita}},\ }\href {\doibase 10.1088/1475-7516/2017/01/019} {\bibfield
  {journal} {\bibinfo  {journal} {JCAP}\ }\textbf {\bibinfo {volume} {01}},\
  \bibinfo {pages} {019} (\bibinfo {year} {2017})},\ \Eprint
  {http://arxiv.org/abs/1608.04216} {arXiv:1608.04216 [astro-ph.CO]}
  \BibitemShut {NoStop}%
\bibitem [{\citenamefont {Maleknejad}\ and\ \citenamefont
  {Sheikh-Jabbari}(2013)}]{Maleknejad:2011jw}%
  \BibitemOpen
  \bibfield  {author} {\bibinfo {author} {\bibfnamefont {A.}~\bibnamefont
  {Maleknejad}}\ and\ \bibinfo {author} {\bibfnamefont {M.~M.}\ \bibnamefont
  {Sheikh-Jabbari}},\ }\href {\doibase 10.1016/j.physletb.2013.05.001}
  {\bibfield  {journal} {\bibinfo  {journal} {Phys. Lett. B}\ }\textbf
  {\bibinfo {volume} {723}},\ \bibinfo {pages} {224} (\bibinfo {year}
  {2013})},\ \Eprint {http://arxiv.org/abs/1102.1513} {arXiv:1102.1513
  [hep-ph]} \BibitemShut {NoStop}%
\bibitem [{\citenamefont {Maleknejad}\ and\ \citenamefont
  {Sheikh-Jabbari}(2011)}]{Maleknejad:2011sq}%
  \BibitemOpen
  \bibfield  {author} {\bibinfo {author} {\bibfnamefont {A.}~\bibnamefont
  {Maleknejad}}\ and\ \bibinfo {author} {\bibfnamefont {M.~M.}\ \bibnamefont
  {Sheikh-Jabbari}},\ }\href {\doibase 10.1103/PhysRevD.84.043515} {\bibfield
  {journal} {\bibinfo  {journal} {Phys. Rev. D}\ }\textbf {\bibinfo {volume}
  {84}},\ \bibinfo {pages} {043515} (\bibinfo {year} {2011})},\ \Eprint
  {http://arxiv.org/abs/1102.1932} {arXiv:1102.1932 [hep-ph]} \BibitemShut
  {NoStop}%
\bibitem [{\citenamefont {Fujita}\ \emph {et~al.}(2018)\citenamefont {Fujita},
  \citenamefont {Namba},\ and\ \citenamefont {Tada}}]{Fujita:2017jwq}%
  \BibitemOpen
  \bibfield  {author} {\bibinfo {author} {\bibfnamefont {T.}~\bibnamefont
  {Fujita}}, \bibinfo {author} {\bibfnamefont {R.}~\bibnamefont {Namba}}, \
  and\ \bibinfo {author} {\bibfnamefont {Y.}~\bibnamefont {Tada}},\ }\href
  {\doibase 10.1016/j.physletb.2017.12.014} {\bibfield  {journal} {\bibinfo
  {journal} {Phys. Lett. B}\ }\textbf {\bibinfo {volume} {778}},\ \bibinfo
  {pages} {17} (\bibinfo {year} {2018})},\ \Eprint
  {http://arxiv.org/abs/1705.01533} {arXiv:1705.01533 [astro-ph.CO]}
  \BibitemShut {NoStop}%
\bibitem [{\citenamefont {Papageorgiou}\ \emph {et~al.}(2019)\citenamefont
  {Papageorgiou}, \citenamefont {Peloso},\ and\ \citenamefont
  {Unal}}]{Papageorgiou:2019ecb}%
  \BibitemOpen
  \bibfield  {author} {\bibinfo {author} {\bibfnamefont {A.}~\bibnamefont
  {Papageorgiou}}, \bibinfo {author} {\bibfnamefont {M.}~\bibnamefont
  {Peloso}}, \ and\ \bibinfo {author} {\bibfnamefont {C.}~\bibnamefont
  {Unal}},\ }\href {\doibase 10.1088/1475-7516/2019/07/004} {\bibfield
  {journal} {\bibinfo  {journal} {JCAP}\ }\textbf {\bibinfo {volume} {07}},\
  \bibinfo {pages} {004} (\bibinfo {year} {2019})},\ \Eprint
  {http://arxiv.org/abs/1904.01488} {arXiv:1904.01488 [astro-ph.CO]}
  \BibitemShut {NoStop}%
\bibitem [{\citenamefont {Dimastrogiovanni}\ and\ \citenamefont
  {Peloso}(2013)}]{Dimastrogiovanni:2012ew}%
  \BibitemOpen
  \bibfield  {author} {\bibinfo {author} {\bibfnamefont {E.}~\bibnamefont
  {Dimastrogiovanni}}\ and\ \bibinfo {author} {\bibfnamefont {M.}~\bibnamefont
  {Peloso}},\ }\href {\doibase 10.1103/PhysRevD.87.103501} {\bibfield
  {journal} {\bibinfo  {journal} {Phys. Rev. D}\ }\textbf {\bibinfo {volume}
  {87}},\ \bibinfo {pages} {103501} (\bibinfo {year} {2013})},\ \Eprint
  {http://arxiv.org/abs/1212.5184} {arXiv:1212.5184 [astro-ph.CO]} \BibitemShut
  {NoStop}%
\bibitem [{\citenamefont {Mangano}\ \emph {et~al.}(2005)\citenamefont
  {Mangano}, \citenamefont {Miele}, \citenamefont {Pastor}, \citenamefont
  {Pinto}, \citenamefont {Pisanti},\ and\ \citenamefont
  {Serpico}}]{Mangano:2005cc}%
  \BibitemOpen
  \bibfield  {author} {\bibinfo {author} {\bibfnamefont {G.}~\bibnamefont
  {Mangano}}, \bibinfo {author} {\bibfnamefont {G.}~\bibnamefont {Miele}},
  \bibinfo {author} {\bibfnamefont {S.}~\bibnamefont {Pastor}}, \bibinfo
  {author} {\bibfnamefont {T.}~\bibnamefont {Pinto}}, \bibinfo {author}
  {\bibfnamefont {O.}~\bibnamefont {Pisanti}}, \ and\ \bibinfo {author}
  {\bibfnamefont {P.~D.}\ \bibnamefont {Serpico}},\ }\href {\doibase
  10.1016/j.nuclphysb.2005.09.041} {\bibfield  {journal} {\bibinfo  {journal}
  {Nucl. Phys. B}\ }\textbf {\bibinfo {volume} {729}},\ \bibinfo {pages} {221}
  (\bibinfo {year} {2005})},\ \Eprint {http://arxiv.org/abs/hep-ph/0506164}
  {arXiv:hep-ph/0506164} \BibitemShut {NoStop}%
\bibitem [{\citenamefont {Grohs}\ \emph {et~al.}(2016)\citenamefont {Grohs},
  \citenamefont {Fuller}, \citenamefont {Kishimoto}, \citenamefont {Paris},\
  and\ \citenamefont {Vlasenko}}]{Grohs:2015tfy}%
  \BibitemOpen
  \bibfield  {author} {\bibinfo {author} {\bibfnamefont {E.}~\bibnamefont
  {Grohs}}, \bibinfo {author} {\bibfnamefont {G.~M.}\ \bibnamefont {Fuller}},
  \bibinfo {author} {\bibfnamefont {C.~T.}\ \bibnamefont {Kishimoto}}, \bibinfo
  {author} {\bibfnamefont {M.~W.}\ \bibnamefont {Paris}}, \ and\ \bibinfo
  {author} {\bibfnamefont {A.}~\bibnamefont {Vlasenko}},\ }\href {\doibase
  10.1103/PhysRevD.93.083522} {\bibfield  {journal} {\bibinfo  {journal} {Phys.
  Rev. D}\ }\textbf {\bibinfo {volume} {93}},\ \bibinfo {pages} {083522}
  (\bibinfo {year} {2016})},\ \Eprint {http://arxiv.org/abs/1512.02205}
  {arXiv:1512.02205 [astro-ph.CO]} \BibitemShut {NoStop}%
\bibitem [{\citenamefont {de~Salas}\ and\ \citenamefont
  {Pastor}(2016)}]{deSalas:2016ztq}%
  \BibitemOpen
  \bibfield  {author} {\bibinfo {author} {\bibfnamefont {P.~F.}\ \bibnamefont
  {de~Salas}}\ and\ \bibinfo {author} {\bibfnamefont {S.}~\bibnamefont
  {Pastor}},\ }\href {\doibase 10.1088/1475-7516/2016/07/051} {\bibfield
  {journal} {\bibinfo  {journal} {JCAP}\ }\textbf {\bibinfo {volume} {07}},\
  \bibinfo {pages} {051} (\bibinfo {year} {2016})},\ \Eprint
  {http://arxiv.org/abs/1606.06986} {arXiv:1606.06986 [hep-ph]} \BibitemShut
  {NoStop}%
\bibitem [{\citenamefont {Maleknejad}\ and\ \citenamefont
  {Komatsu}(2019)}]{Maleknejad:2018nxz}%
  \BibitemOpen
  \bibfield  {author} {\bibinfo {author} {\bibfnamefont {A.}~\bibnamefont
  {Maleknejad}}\ and\ \bibinfo {author} {\bibfnamefont {E.}~\bibnamefont
  {Komatsu}},\ }\href {\doibase 10.1007/JHEP05(2019)174} {\bibfield  {journal}
  {\bibinfo  {journal} {JHEP}\ }\textbf {\bibinfo {volume} {05}},\ \bibinfo
  {pages} {174} (\bibinfo {year} {2019})},\ \Eprint
  {http://arxiv.org/abs/1808.09076} {arXiv:1808.09076 [hep-ph]} \BibitemShut
  {NoStop}%
\bibitem [{\citenamefont {Ade}\ \emph {et~al.}(2016)\citenamefont {Ade} \emph
  {et~al.}}]{Planck:2015zfm}%
  \BibitemOpen
  \bibfield  {author} {\bibinfo {author} {\bibfnamefont {P.~A.~R.}\
  \bibnamefont {Ade}} \emph {et~al.} (\bibinfo {collaboration} {Planck}),\
  }\href {\doibase 10.1051/0004-6361/201525836} {\bibfield  {journal} {\bibinfo
   {journal} {Astron. Astrophys.}\ }\textbf {\bibinfo {volume} {594}},\
  \bibinfo {pages} {A17} (\bibinfo {year} {2016})},\ \Eprint
  {http://arxiv.org/abs/1502.01592} {arXiv:1502.01592 [astro-ph.CO]}
  \BibitemShut {NoStop}%
\bibitem [{\citenamefont {Aghanim}\ \emph {et~al.}(2020)\citenamefont {Aghanim}
  \emph {et~al.}}]{Planck:2018vyg}%
  \BibitemOpen
  \bibfield  {author} {\bibinfo {author} {\bibfnamefont {N.}~\bibnamefont
  {Aghanim}} \emph {et~al.} (\bibinfo {collaboration} {Planck}),\ }\href
  {\doibase 10.1051/0004-6361/201833910} {\bibfield  {journal} {\bibinfo
  {journal} {Astron. Astrophys.}\ }\textbf {\bibinfo {volume} {641}},\ \bibinfo
  {pages} {A6} (\bibinfo {year} {2020})},\ \Eprint
  {http://arxiv.org/abs/1807.06209} {arXiv:1807.06209 [astro-ph.CO]}
  \BibitemShut {NoStop}%
\end{thebibliography}%

\end{document}